\documentclass[preprint]{./acm_proc_article-sp}
\usepackage{epsfig}
\usepackage[square,comma,numbers,sort&compress]{natbib}

\newcommand{\UPLB}{University of the Philippines Los Ba\~{n}os}

\begin{document}
\title{Simulating the Effects of\\Various Road Infrastructure Improvements to\\Vehicular Traffic in a Busy Three-road Fork}
\numberofauthors{3}
\author{
\alignauthor Marian G. Arada\\
   \affaddr{Polytechnic University of the Philippines-Taguig\\Taguig City 1632, Metro Manila}\\
\alignauthor Merly F. Tataro\\
   \affaddr{Polytechnic University of the Philippines-Taguig\\Taguig City 1632, Metro Manila}\\
\alignauthor Jaderick P. Pabico\\
   \affaddr{Institute of Computer Science}\\
   \affaddr{\UPLB\\College 4031, Laguna}
}
\date{}
\maketitle

\begin{abstract}
Using microsimulations of vehicular dynamics, we studied the effects of several proposed infrastructure developments to the mean travel delay time~$\Delta$ and mean speed~$\Sigma$ of vehicles passing a busy three-road fork, particularly in the non-signalized roundabout junction of Lower Bicutan, Taguig City, Metro Manila.  We designed and implemented multi-agent-based microsimulation models to mimic the autonomous driving behavior of heterogeneous individuals and measured the effect of various proposed infrastructure developments on~$\Delta$ and~$\Sigma$. Our aim is to find out the best infrastructure development from among three choices being considered by the local government for the purpose of solving the traffic problems in the area. We created simulation models of the current vehicular traffic situation in the area using the mean travel times~$\tau$ of statistically sampled vehicles to show that our model can simulate the real-world at a significance level of $\alpha=0.05$.  Based on these models, we then simulated the effect of the proposed infrastructure developments on~$\Delta$ and~$\Sigma$ and used these metrics as our basis of comparison.  We found out that the proposed widening of one fork from two lanes to three lanes has the most improved metrics at the same $\alpha=0.05$ compared to the metrics we observed in the current situation. Under this infrastructure development, the~$\Delta$ increases linearly ($R^2=0.98$) at the rate of 1.03~$s$, while the~$\Sigma$ decreases linearly ($R^2>0.99$) at the rate of 0.14~$km/h$ per percent increase in the total vehicle volume~$\mathcal{V}$.

\category{K.3.7.1}{Computing Methodologies}{Artificial Intelligence}[Multi-agent systems]
\category{K.5.1.2}{Computing Methodologies}{Modeling and Simulation}[Model verification and validation]
\category{L.9.5}{Applied Computing}{Operations Research}[Transportation]

\end{abstract}

\section{Introduction}
\begin{figure*}[bt]
\centering\epsfig{file=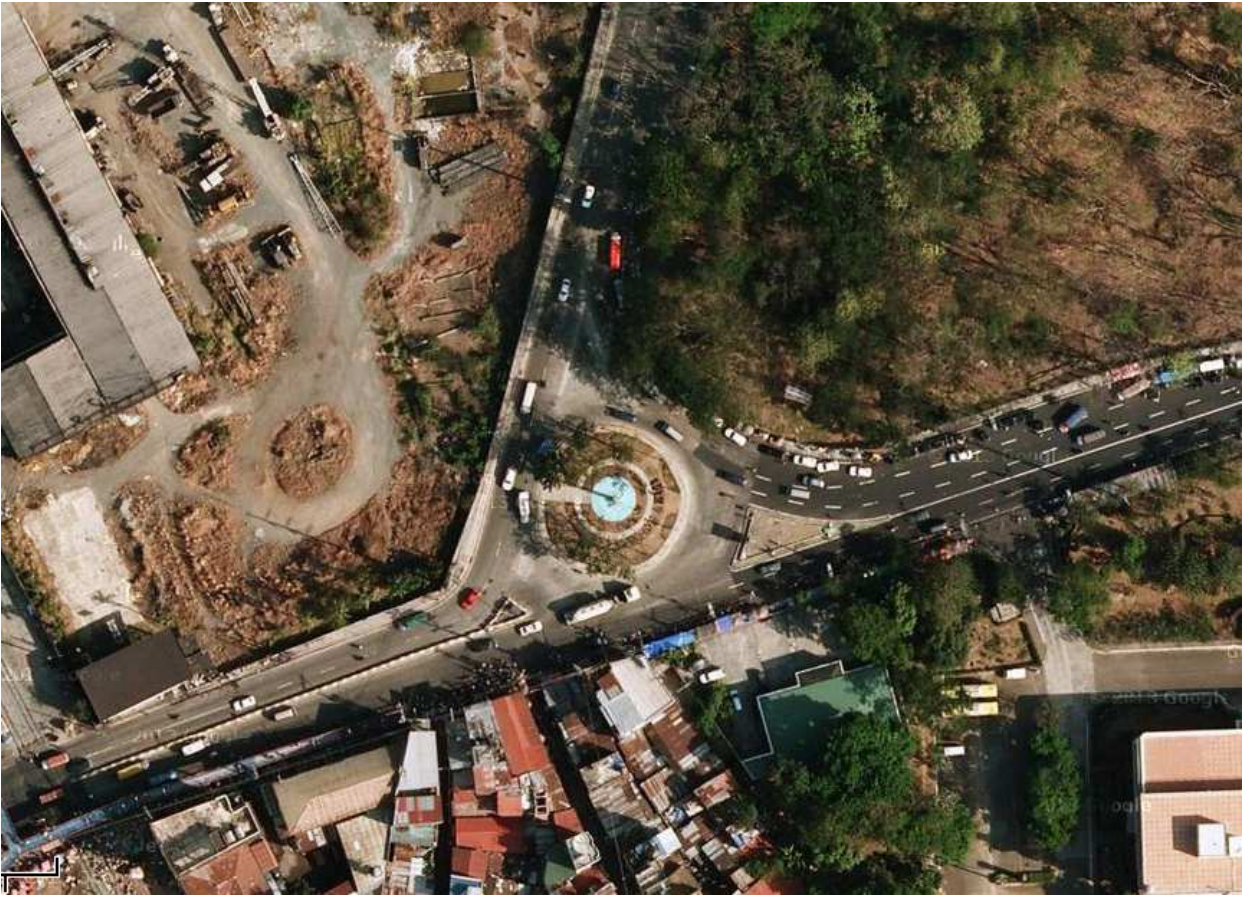, width=6.5in}
\caption{The area map of Upper Bicutan, Taguig City showing the relative location of BR~\citep{googlemap13}.}\label{fig:BR-map}
\end{figure*}

The increasing number of vehicles that use the road at the same time is causing congestions on road networks  especially at road sections where nominal widths change such as intersections, three-way forks, and roundabouts. One real-world example of such width change is the Bicutan Roundabout (BR) in Taguig City, Metro Manila. Commuters passing through this roundabout experience delay in travel everyday, especially during rush hours on week days, because of vehicular congestion in the area. BR is located at the Upper Bicutan area and has an inscribed diameter of 34~$m$. It is bounded to the East by the Department of Science and Technology (DOST) campus, to the North by the Philippine National Construction Corporation (PNCC) campus, to the West by the Philippine National Railways (PNR) crossing, and to the South by North {\em Daanghari}. BR serves as the T-intersection between General Paulino Santos Avenue (GPSA) which lies along the ENE-WSW line and SLEX's East Service Road (ESR) which lies along the NEN-SWS line of the area (Figure~\ref{fig:BR-map}). The nominal center of BR is approximately 100~$m$ WSW from DOST's main gate along GPSA, 150~$m$ SWS of the ESR and San Martin de Porres fork, and 106~$m$ ENE of PNR crossing. Jeepneys, buses, trucks, taxis, AUVs, motorcycles, tricycles, and bicycles are the usual type of vehicles that pass the area. Upper Bicutan, a barangay under the urbanized City of Taguig, is situated on the western shore of Laguna de Bay and is bordered in the south by Muntinlupa City, in the southwest by Para\~naque City, in the west by Cainta, Rizal, in the northeast by Taytay, Rizal and in the north by Makati City, Pasig City and the Municipality of Pateros. According to the 2010 census, 62,570 out of the city's total population of 613,343 are residents of Upper Bicutan~\citep{taguig-web}.

The thrust of the city government administration is to make Taguig City a ``Premier City'' in the Philippines. However,  the physical and financial development of a city is greatly dependent, among others, on the efficiency of its transportation support system, particularly how the city manages the vehicular flow along its road networks. The efficient flow of vehicles carrying human and non-human capitals results in the efficient flow of resources. This translates to a productive citizenry and successful businesses. Because of this, the city government must {\em make its move} in pursuing efficient and practical steps to at least mitigate, if not totally eradicate, traffic congestion which must result into improved vehicular flow. Vehicular flow may be quantified by the following metrics and we say that improvement has been achieved if there is a statistically significant difference the measurement at a particular significance level~$\alpha$ as compared to the current situation:
\begin{enumerate}
\item Average vehicular delay time while in BR~($\Delta$) -- This metric measures the total amount of time that the vehicle is in the stopped position (i.e., when the vehicle has a speed of 0~$km/h$); and
\item Average vehicular speed while in BR~($\Sigma$) -- This metric measures the mean speed of a vehicle while traveling within the BR area. 
\end{enumerate}
Ideally, we wanted~$\Delta$ to approximate zero while we wanted~$\Sigma$ to be near but not more than the legal maximum speed. In realistic situations, however, $\Delta>0$ and $\Sigma$~is even way below the legal minimum speed. These two metrics complement each other because for some agents such as those who drive big trucks, an almost zero~$\Delta$ and a very low~$\Sigma$ is much more desirable than a $\Delta>0$ and a relatively high~$\Sigma$. This is because it requires much more fuel to move the heavy vehicles from a full stop than to keep them moving at low~$\Sigma$. For agent drivers of light- and medium-weight vehicles, a $\Delta>0$ but a very high~$\Sigma$ is desirable because fuel burnt at low~$\Sigma$ is greater than when idle (for at least within some reasonable time).

There are two ways the city can {\em make its move}: (1)~Improve the infrastructure of the road network that will best optimize the~$\Delta$ and the~$\Sigma$, or (2)~implement responsive but efficient time-dependent traffic schemes that will significantly improve both metrics. Both moves are very costly if the government is to try all possible schemes and infrastructure just for the simple reason of finding the optimal one. 

In terms of finding the optimal infrastructure improvement for BR, one should not do it via a trial-and-error way because it means the following: (1)~building the infrastructure, (2)~letting it tried by the commuters for a reasonable amount of time just to statistically compile the~$\Delta$ and the~$\Sigma$ metrics, and then (3)~removing it to build the next one. By the time the government has compiled the data it needed, it has already spent significant amount of money, time, and other resources. Besides, the data collected might have already been biased due to what we call ``data collection dependency'' brought about by the memory of the agents in their experiences. This means that for a trial-and-error with $n$~iterations, the data collected on the $(i+1)$th iterate might have been interactively affected by the improvement made during the $i$th iterate, unless of course, after every iterate we revert back to the current infrastructure for the same amount of time just so both the drivers and the commuters will forget their $i$th~experiences. 

In terms of implementing responsive yet efficient traffic schemes, most local governments have the habitual liking of doing it in the trial-and-error way, as experienced by all of this paper's authors. If for example the local Traffic Management Office (TMO) is considering $n$~traffic schemes, they will: (1)~implement the $i$th scheme for a $t$~period of time (where $t$~is empirically selected to be long enough for the drivers and the commuters to adapt to it), (2)~compile the~$\Delta$ and the~$\Sigma$ metrics, and (3)~repeat the process for the $(i+1)$th scheme. The local TMO will then compare the schemes and decide based on the metrics collected without considering (maybe even knowing) that there is an interactive effect on the $(i+1)$th scheme the experiences and adaptation of the driver and commuter agents during the implementation of the $i$th scheme.

One efficient way that may be used to try out ideas and schemes in solving traffic congestions is to use computer-based simulation, a model created to reflect real-world situations. Once the model is constructed, solutions can be tried out to determine its impact on traffic congestion~\citep{bared02}, without the need for a costly real-world trial-and-error implementation. In this study, we used a multi-agent framework to simulate the object-following behavior of heterogeneous agents that share a road network. The road network reflects the scaled-down dimension of a real-world network, such as that of the BR, servicing scaled-down dimensions of real-world vehicles. Our model can be characterized as microscopic, stochastic, discrete time-step, and behavior-based. The model is microscopic because it simulates the behavior of the vehicles themselves while on the road in the presence of, and interacting with, other vehicles. The model is stochastic because it is based on random events that follow some distributions that we observed in the real-world. The model is discrete time-step because at each unit time, the agents concurrently change their respective positions, directions, and speeds driven only by their respective simulated behaviors in response to their respective interactions with the current state of the road network. The model is behavior-based because the traffic flow algorithms that we used are based on a psycho-physical car following model for vehicles moving along the length of the road~\citep{fritzsche94,gipps81}. We used a rule-based algorithm for lateral movement that simulates the lane-changing behaviors of the agents~\citep{alshihabi03,VISSIM90}. 

We present in this paper the result of our experiments in searching for the optimal infrastructure improvement for BR from among three choices that are being considered by the local government. We  present our results in searching for the optimal traffic schemes for the current infrastructure in another paper~\citep{tataro13}. Our study had three objectives: (1)~To show that our model reflects the current real-world scenario at a statistical significance of $\alpha=0.05$ using the mean travel time~$\tau$ of the sampled vehicles as metric, (2)~to use the model to simulate the effect on~$\Delta$ and on~$\Sigma$ of three infrastructure developments (ID) in BR currently being considered by the local government, and (3)~simulate the effect on~$\Delta$ and on~$\Sigma$ the 10\%, 50\%, and 100\% increase in vehicular volume under the best ID. Based on our results, we found out that: (1)~The average travel time of the simulated vehicles while at the BR is not significantly different at $\alpha=0.05$ from the observed average travel time of the sampled real-world vehicles, (2)~the best ID is when the South bound lane of the ESR is widen up to three lanes continuing to the west bound lane of GPSA from BR to PNR, and~(3) the~$\Delta$ increases linearly ($R^2=0.98$) at the rate of 1.03~$s$ per percent increase in the total vehicle volume~$\mathcal{V}$, while the~$\Sigma$ decreases linearly ($R^2>0.99$) at the rate of 0.14~$km/h$ per percent increase in~$\mathcal{V}$.

\section{Model}

A computer model that controls an agent's behavior (i.e., the driver) in terms of its reaction to the vehicle in front of it while at the same lane is called a {\em car-following model}~\citep{herman63}. An agent-driven vehicle is said to be {\em following} when it is constrained by the speed of the moving vehicle in front of it, and that driving at the agent's desired speed will lead to a collision. When a driver agent is not constrained by another vehicle, it is said to be {\em free} and travels, in general, at its desired speed. The actions of the {\em following} agent is defined by the agent's acceleration, although some models, for example the car-following model developed by~\citet{gipps81}, define the agent's actions through the agent's speed. Some car-following models only describe the agents' behavior when they are currently following another vehicle, while other models determine the agents' behavior in all situations. We believe, however, that to model the real-world decision-making capabilities of human drivers, a car-following model should identify both of the following:
\begin{enumerate}
\item The current state~$S$ the vehicle is in; and
\item What actions~$A$ are desirable at this state.
\end{enumerate}
Most car-following models use several states to describe the following agents' behavior. Most models use:
\begin{enumerate}
\item $S_f$: A state for {\bf free driving}, where the vehicles are unconstrained and the respective driver agents try to achieve their desired speeds (subject of course to pertinent legal speed limits of the road network);
\item $S_n$: A state for {\bf normal following}, where the following agents adjust their respective speeds with respect to the speeds of the vehicles in front of them; and
\item $S_e$: A state for an {\bf emergency deceleration}, where agents try to decelerate to avoid a collision with the vehicle in front.
\end{enumerate} 

Throughout the paper, we used the notations summarized in Table~\ref{tab:notations} to describe the kinematic quantities and model outputs.

\begin{table*}[hbt]
\caption{Alphabetical listing of notations, mathematical variables, and abbreviations used in this paper including their respective descriptions.}\label{tab:notations}
\centering\begin{tabular}{cl}
\hline\hline
Notation & Description\\
\hline
$\delta x$ & Space headway between $x_{n-1}$ and~$x_n$ in~$m$\\
$\delta v$ & difference in speed between $x_{n-1}$ and~$x_n$ in~$m/s$\\
$a_n$ & Acceleration of the $n$th vehicle in~$m/s^2$\\
$L_n$ & length of the $n$th vehicle in~$m$\\
$s_n$ & Effective length of the $n$th vehicle in~$m$\\
$T$   & Reaction time in~$s$\\
$v_n$ & Speed of the $n$th vehicle in~$m/s$\\
$v_n^{\rm desired}$ & Desired speed of the $n$th vehicle in~$m/s$\\
$x_n$ & Longitudinal position of the $n$th vehicle in~$m$\\
\hline
$\alpha$ & Statistical level of significance\\
$\Delta$ & Mean delay time per vehicle in~$s$\\
$\Sigma$ & Mean vehicular speed in~$m/s$\\
$\tau$   & Mean vehicular travel time~$s$\\
$\tau_o$ & $\tau$ of the sampled vehicles from observation\\
$\tau_s$ & Mean $\tau$ of the simulated vehicles\\
$\mathcal{V}$ & Total volume of vehicles\\
\hline
ANOVA    & Analysis of Variance using the F-Statistics\\
BR      & Bicutan Roundabout\\
DOST    & Department of Science and Technology\\
ESR     & East Service Road\\
GPSA    & General Paulino Santos Avenue\\
ID      & Infrastructure Development\\
MMDA    & Metro Manila Development Authority\\
PNCC    & Philippine National Construction Corporation\\
PNR     & Philippine National Railways\\
PUP-T   & Polytechnic University of the Philippines-Taguig\\
SLEX    & Southern Luzon Expressway\\
TUP-T   & Technological University of the Philippines-Taguig\\
TMO     & Traffic Management Office\\
\hline\hline
\end{tabular}
\end{table*}

\subsection{$S_f$ Free Driving State}

The agents at the $S_f$ state try to accelerate or decelerate to achieve its current desired speed. If the current speed~$v_i$ of the $i$th vehicle is higher than its desired speed~$v_i^{\rm desired}$, the agents uses the normal deceleration rate ($-a_i^{\rm normal}$) to slow down to the desired speed. If  $v_i < v_i^{\rm desired}$, the agents use its maximum acceleration rate $a_i^{\rm max}$ to reach $v_i^{\rm desired}$ at the shortest time possible. The $-a_i^{\rm normal}$ and $a_i^{\rm max}$ are parameters of the simulation and they are dependent on the type of vehicle the agent is driving. The acceleration rate of the $i$th agent is expressed as
$$
  a_i = \left\{\begin{array}{ll} a_i^{\rm max} & \quad v_i < v_i^{\rm desired}\\
                                  0           & \quad v_i = v_i^{\rm desired}\\
                                a_i^{\rm normal} &  \quad v_i > v_i^{\rm desired}\end{array} \right.
$$
The time~$t_i$ it will take for the $i$th vehicle to achieve its desired speed is
$$
  t_i = \left\{\begin{array}{ll}
                  \frac{v_i}{a_i^{\rm max}} & \quad v_i < v_i^{\rm desired}\\
                  \frac{v_i}{a_i^{\rm normal}} & \quad v_i > v_i^{\rm desired}
               \end{array}\right.
$$
\subsection{$S_n$ Normal Following State}

While at the normal following state $S_n$, the acceleration rate of the $i$th vehicle is given by an asymmetrical Gazis-Herman-Rothery (GHR) model~\citep{chandler58,gazis61,yang96}. The acceleration is computed as
$$
    a_i = r^{\pm}\frac{v_i^{s^{\pm}}}{(x_{i-1}-L_{i-1}-x_i)^{t^{\pm}}}(v_{i-1}-v_i),
$$
where $r^{\pm}$, $s^{\pm}$, and $t^{\pm}$ are model parameters. The parameters $r^+$, $s^+$, and $t^+$ are used if $v_i\le v_{i-1}$, while the parameters $r^-$, $s^-$, and $t^-$ are used if $v_i > v_{i-1}$. Notice here that the final acceleration of the $i$th vehicle is given as $\max\{a_i, a_{i-1}\}$.

\subsection{$S_e$ Emergency State}

Under this state, the agents use a deceleration rate that prevents collision and extends~$\delta x$. This deceleration rate is given by
$$
  a_i = \left\{ \begin{array}{ll}
            \min\{a_i^{\rm normal}, a_{i-1}-\frac{0.5(v_i-v_{i-1})^2}{x_{i-1}-L_{i-1}-x_i}\} & \quad v_i > v_{i-1}\\
            \min\{a_i^{\rm normal}, a_{i-1}+0.25a_i^{\rm normal} & \quad v_i \le v_{i-1}
        \end{array}\right.
$$

\section{Methodology}\label{sec:method}

We present in this section the activities that we performed in this study. We also present here the metrics that we measured, as well as the statistical analysis that we employed to analyze the results of the study.

\pagebreak
\subsection{Interview with Stakeholders}

We conducted interviews in order for us to understand the current plans of the two local governments that have jurisdictions over the BR: the Taguig City Traffic Management Office (TMO) and the Metro Manila Development Authority (MMDA). The interview served as some sort of leveling off with the intended stakeholders. We also wanted the result of this research endeavour to be put to use, particularly because two of the three authors of this paper pass this area on a daily basis. Thus, if either or both local governments will implement the result of our research, we will greatly benefit from such. We also asked permission from the two governments to allow us to conduct scientific observations so that we can quantify the current vehicular flow in BR.


\subsection{Observation of Current Situation}

We employed the help of twelve enumerators whom we assigned to different identified points within the BR junction.  Each point of entrance to and exit from BR has designated persons to record the time of entrance and exit of the vehicle, and the type of vehicle that passed through. Each vehicle is identified by their tag or plate number. We recorded a series of actual observations during peak and non-peak periods for us to be able to get a statistically accurate data and determine the highest and lowest traffic volume.  From these observations, we were able to obtain the mean travel time~$\tau$ of sampled vehicles, as well as the respective distributions of each vehicle type. To identify the specific points in the BR, we divided the area into six routes as follows:
\begin{enumerate}
\item {\bf Route 1} is the route from PNR to DOST Campus Eastbound along the GPSA passing through the BR;
\item {\bf Route 2} is the route from PNR to PNCC Campus, Eastbound along GPSA from the PNR to BR, and then turning left through the BR, and up to the Northbound lane of ESR going towards the PNCC Campus;
\item {\bf Route 3} is the route from DOST campus to PNR Westbound along the GPSA passing through the BR;
\item {\bf Route 4} is the route from the DOST Campus to the PNR Crossing, Westbound along the GPSA up to the BR, and then turning right through the BR, and up to the Northbound lane of ESR going towards the PNCC Campus;
\item {\bf Route 5} is the route from PNCC Campus to PNR, Southbound along the ESR, turning right through BR, and then Westbound along the GPSA to the PNR; and
\item {\bf Route 6} is the route from PNCC Campus to the DOST Campus, Southbound along the ESR, turning left through BR, and then Eastbound along the GPSA to the DOST Campus.
\end{enumerate}

\subsection{Simulation of Current Situation}

We conducted a replicated microsimulation study of the vehicular flow under the current BR using the data on respective distribution of vehicles by type. We replicated the study to $n=10$ and computed the mean~$\tau$. We wanted to know if the model can statistically reproduce the observed vehicular traffic data. Statistically, the respective differences of the~$\tau$ between the observed ($\tau_o$) and the mean simulated ($\tau_s$) runs must not be different from zero at $\alpha=0.05$. We used analysis of variance statistics (ANOVA) to evaluate two hypotheses, the null hypothesis $H_0$ and the alternative hypothesis $H_a$ as follows:

$H_0$: There is no significant difference between~$\tau_o$ and~$\tau_s$ at $\alpha=0.05$.

$H_a$: There is significant difference between~$\tau_o$ and~$\tau_s$ at the same $\alpha$.

\subsection{Microsimulations of Proposed Developments}

We created the respective replicated $n=10$ microsimulation studies of the vehicular flow using the observed current vehicular distribution when under the different planned infrastructure developments (ID) of the BR. These IDs are based on the C6 Project blueprint, the medium-term development plan by the City Mayor of Taguig Lani Cayetano, and DOST's Monorail Project which is currently underway. The proposed developments are:
\begin{enumerate}
\item Infrastructure Development 1 (ID1): Widening of the GPSA into three lanes but only those lanes towards the BR. That is the east bound lane from PNR to BR and the west bound lane from DOST to BR.
\item Infrastructure Development 2 (ID2): Widening of the GPSA into three lanes but only the eastbound lane from PNR to BR continuing from BR to DOST gate.
\item Infrastructure Development 3 (ID3): Widening of the south bound lane of the ESR into three lanes up to the BR continuing to the west bound lane of GPSA from BR to PNR.
\end{enumerate}
We computed the respective means of~$\Delta$ and~$\Sigma$ for all infrastructure improvements, including that of the current infrastructure we termed here as ID0 (i.e., from ID0 through ID3) and conducted an analysis of variance (ANOVA) statistics to find out whether the respective means are significantly different from each other at $\alpha=0.05$. That is, we want to test two hypotheses, the null hypothesis $H_0$ and the alternative hypothesis $H_a$ as follows:

$H_0$: The absolute pairwise differences  $|\Delta_{\rm ID0} - \Delta_{\rm ID1}| = \dots = |\Delta_{\rm ID1} - \Delta_{\rm ID2}| = |\Delta_{\rm ID1} - \Delta_{\rm ID3}| = |\Delta_{\rm ID2} - \Delta_{\rm ID3}| = 0$, where the $=$ sign in this sentence would mean ``not significantly different from'' or ``statistically equal to.''

$H_a$: Any of the following respective pairwise absolute differences is true: $|\Delta_{\rm ID0} - \Delta_{\rm ID1}| > 0$, $\dots$ , $|\Delta_{\rm ID1} - \Delta_{\rm ID2}| > 0$, $|\Delta_{\rm ID1} - \Delta_{\rm ID3}| > 0$,  or $|\Delta_{\rm ID2} - \Delta_{\rm ID3}| > 0$.

We have the same set of hypotheses for $\Sigma$.

\subsection{Increased Traffic Volume at Best ID}

We rerun the respective replicated microsimulation studies under ID0, ID1, ID2, and ID3 but with respective increases of 10\%, 50\%, and 100\% in~$\mathcal{V}$, making sure that the distribution by vehicular type is preserved. Here, we wanted to know if the benefits of the proposed improvements will be carried with the increase in~$\mathcal{V}$ and if so, up to how much increase. The assumed increase in~$\mathcal{V}$  is just a natural reaction of the driving agents when there is a perceive improvement in the current situation. The improvement of the vehicular flow due to infrastructure improvements will attract more agents, increasing~$\mathcal{V}$ in the area, and thereby might degrade the expected designed benefits of the improvement in the long run.

\section{Results and Discussion}\label{sec:results}

\subsection{Actual Current Vehicular Flow}

Tables~\ref{tab:current} and~\ref{tab:vehicle-type} summarize the distribution of vehicle per route and per vehicle type, respectively, that we compiled during the actual observation. Route~4 has the most number of total vehicles that passed during the observation period with 23.79\%. This is followed by Routes~5, 1, and~2, in non-increasing order. Route 4, undeniably has the most number of vehicles since Route~4 is the only route that may be used by residents from the Lower Bicutan area who are trying to go to the Metropolitan Manila Area via the C-5 Diversion. Aside from the Lower Bicutan residents, the route is also extensively used by students and employees of several Higher Education Institutions (HEIs) and government agencies such as the Technological University of the Philippines-Taguig (TUP-T), the Polytechnic University of the Philippines-Taguig (PUP-T), the LGU-funded Taguig City University, and DOST's national offices and laboratories. Route~5 is extensively used by trucking companies with medium-length trucks carrying heavy cargoes. These vehicles are coming off from the C-5 Diversion going to the Southern Luzon Area via the SLEX. Route~1 is obviously used by commuters coming from the SLEX area going to the residential Lower Bicutan or to any of the HEIs and government agencies. Route~2 is the reverse of Route~5, where mostly medium-length trucks pass through carrying cargoes from the Southern Luzon area via SLEX going to Metropolitan Manila Area through the C-5 Diversion. From Table~\ref{tab:vehicle-type}, motorcycles and cars combined account for about 70\% of the vehicles that use BR. This is followed by bicycles, jeepneys and vans. The heavier vehicles, such as the many-wheelers and buses only account for about 4\% combined. This distribution by type is actually expected because the area is used extensively by residents and commuters rather than by non-human cargoes. Table~\ref{tab:vehicle-dimension} shows the mean vehicular dimensions by vehicle type. We used these dimensions to scale the respective agents in our simulations.

\begin{table}[bth]
\caption{Distribution of the total number of vehicles that passed through the BR during the observation period.}\label{tab:current}
\centering\begin{tabular}{lrr}
\hline\hline
Route & Number of Vehicles & Percentage (\%) \\
\hline
{\bf Route 1} & 778 & 20 \\
{\bf Route 2} & 719 & 18 \\
{\bf Route 3} & 524 & 13 \\
{\bf Route 4} & 934 & 24 \\
{\bf Route 5} & 815 & 21 \\
{\bf Route 6} & 156 & 4 \\
\hline
{\bf Total} & 3,926 \\
\hline\hline
\end{tabular}
\end{table}

\begin{table}[bth]
\caption{Distribution of the vehicles per type.}\label{tab:vehicle-type}
\centering\begin{tabular}{lr}
\hline\hline
Type &  Percentage (\%) \\
\hline
Motorcycle           &  38.30\\
$4\times 8$ wheeler  &   0.64\\
$4\times 6$ wheeler  &   2.75\\
Van                  &   5.88\\
Jeepney              &   9.60\\
Car                  &  31.73\\
Bus                  &   0.38\\
Bicycle              &  10.72\\
\hline
Total                & 100.00\\
\hline\hline
\end{tabular}
\end{table}

\begin{table}[bth]
\caption{The average dimension of observed vehicles per type.}\label{tab:vehicle-dimension}
\centering\begin{tabular}{lrr}
\hline\hline
Vehicle & Length & Width \\
Type    & ($m$)    & ($m$)  \\
\hline
Motorcycle           &  2.00 & 1.5\\
$4\times 8$ wheeler  &  6.59 & 1.5\\
$4\times 6$ wheeler  &  5.41 & 1.5\\
Van                  &  5.50 & 1.5\\
Jeepney              &  4.00 & 1.5\\
Car                  &  4.50 & 1.5\\
Bus                  & 11.54 & 2.5\\
Bicycle              &  1.45 & 0.5\\
\hline\hline
\end{tabular}
\end{table}

\subsection{Evaluation of the Model against the Observed}

Table~\ref{tab:model-validation} summarizes the statistics we computed after conducting ANOVA on~$\tau_o$ and~$\tau_s$. Since we see that the F statistics $\alpha_{\rm F}=0.6873 > \alpha = 0.05$, then we accept the null hypothesis $H_0$ and we say that the difference between the mean sampled observed data $\tau_o$ and the mean simulated data $\tau_s$ is not significantly different from zero at $\alpha \approx 0.05$. Thus, our microsimulation model was able to mimic the driver's behavior in the real world with a guarantee of being correct $1 - \alpha = 0.95$ of the time.

\begin{table*}[bth]
\caption{The ANOVA table comparing $\tau_o$ and $\tau_s$ using the F statistics. SOV means Source of Variation and DF means Degree of Freedom}\label{tab:model-validation}
\centering\begin{tabular}{lrrrrr}
\hline\hline
SOV & DF & Sum of Squares & Mean Square & F & $\alpha_{\rm F}$\\
\hline
Replication            & 9  & 1,633.05 & 181.45 & 39.21 & $<$ 0.0001 \\
$\tau_o$ vs. $\tau_s$  & 1  &     0.80 &   0.80 &  0.17 & 0.6873 \\
Error                  & 9  &    41.65 &   4.63 & \\
\hline
Total                  & 19 & 1,675.50 & \\
\hline\hline
\end{tabular}
\end{table*}

\subsection{Comparison of the IDs}

\begin{figure}[bth]
\centering\epsfig{file=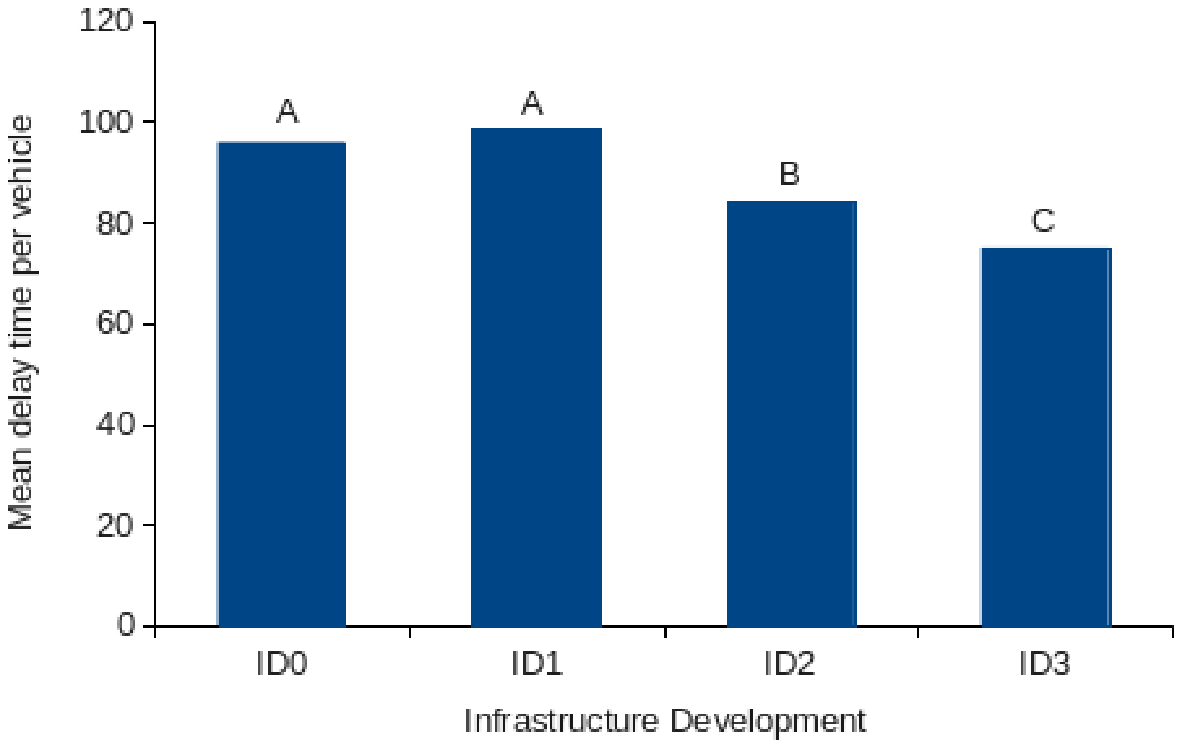, width=3in}
\caption{The respective mean~$\Delta$ of the various IDs with DMRT grouping. Means with the same letter are not statistically different from each other by DMRT.}\label{fig:dmrt1}
\end{figure}

\begin{figure}[bth]
\centering\epsfig{file=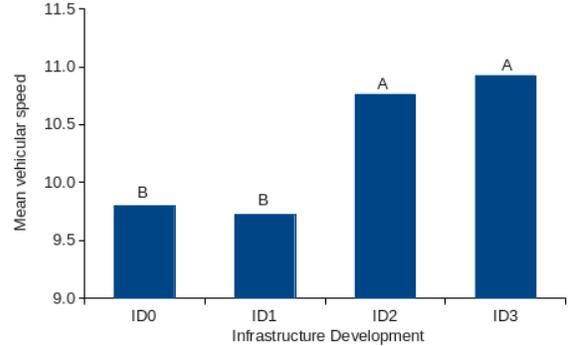, width=3in}
\caption{The respective mean~$\Sigma$ of the various IDs with DMRT grouping. Means with the same letter are not statistically different from each other by DMRT.}\label{fig:dmrt2}
\end{figure}

Tables~\ref{tab:anova-delay} and~\ref{tab:anova-speed} show the respective ANOVA tables of the mean~$\Delta$ and mean~$\Sigma$ for ID0, ID1, ID2, and ID3 where we find that $\alpha_{\rm F}^{\Delta} < \alpha$ and $\alpha_{\rm F}^{\Sigma} < \alpha$. Based on the computed $\alpha_{\rm F}^{\Delta}$ and $\alpha_{\rm F}^{\Sigma}$, we accept the alternate hypothesis $H_a$ saying with confidence $1 - \alpha_{\rm F} = 0.9999$ that at least one of the absolute pairwise difference between $\Delta_{{\rm ID}x}$ and $\Delta_{{\rm ID}y}$, and between $\Sigma_{{\rm ID}x}$ and $\Sigma_{{\rm ID}y}$, $\forall x\ne y$, is significantly greater than zero. Figures~\ref{fig:dmrt1} and~\ref{fig:dmrt2} show the respective mean~$\Delta$ and mean~$\Sigma$ of ID$x$, $\forall x=0,\dots,3$. Based on the DMRT, we see that ID1 will not improve~$\Delta$ nor~$\Sigma$. In fact, both~$\Delta$ and~$\Sigma$ under ID1 is no different, statistically, than the~$\Delta$ and~$\Sigma$ of the current infrastructure, and thus, implementing ID1 will just be a waste of taxpayer's money. We also see that ID2 has significantly improved~$\Delta$ and~$\Sigma$ compared to the current infrastructure. However, ID3 has the most significant improvement in terms of both~$\Delta$ and $\Sigma$ compared to the current infrastructure. From Figure~\ref{fig:dmrt2}, we see that the respective~$\Sigma$ of ID2 and ID3 are not statistically different from each other, so either ID can provide the best~$\Sigma$. Based on the two metrics, however, we say that ID3 can provide the improvement that we desire for the vehicular flow.

\begin{table*}[bth]
\caption{The ANOVA table comparing the different IDs in terms of~$\Delta$ using the F statistics. SOV means Source of Variation and DF means Degree of Freedom}\label{tab:anova-delay}
\centering\begin{tabular}{lrrrrr}
\hline\hline
SOV & DF & Sum of Squares & Mean Square & F & $\alpha_{\rm F}^{\Delta}$\\
\hline
Replication            &  9 & 2,328.63 &   258.74 &  5.12 & 0.0004 \\
ID                     &  3 & 3,556.14 & 1,185.38 & 23.47 & $<$ 0.0001 \\
Error                  & 27 & 1,363.39 &    50.50 & \\
\hline
Total                  & 39 & 7,248.15 & \\
\hline\hline
\end{tabular}
\end{table*}

\begin{table*}[bth]
\caption{The ANOVA table comparing the different IDs in terms of~$\Sigma$ using the F statistics. SOV means Source of Variation and DF means Degree of Freedom}\label{tab:anova-speed}
\centering\begin{tabular}{lrrrrr}
\hline\hline
SOV & DF & Sum of Squares & Mean Square & F & $\alpha_{\rm F}^{\Sigma}$\\
\hline
Replication            &  9 & 14.73 & 1.64 &  5.63 & 0.0002 \\
ID                     &  3 & 11.83 & 3.94 & 13.56 & $<$ 0.0001 \\
Error                  & 27 &  7.85 & 0.29 & \\
\hline
Total                  & 39 & 34.41 & \\
\hline\hline
\end{tabular}
\end{table*}

\subsection{Effect of Increased Traffic Volume}

\begin{figure}[bth]
\centering\epsfig{file=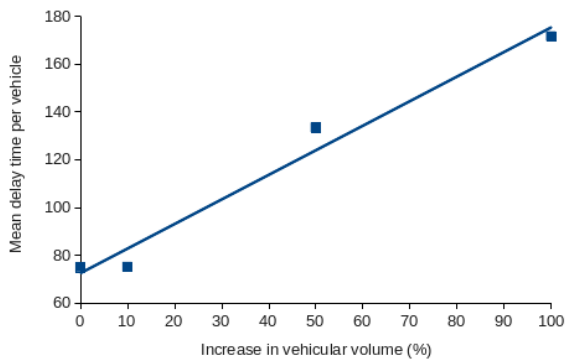, width=3in}
\caption{The respective mean~$\Delta$ of ID3 when~$\mathcal{V}$ was increased from 10\% to 100\%. The line represents the regression with $\Delta = 1.03 \times \mathcal{V} + 72.64$ ($R^2=0.98$).}\label{fig:regress1}
\end{figure}

\begin{figure}[bth]
\centering\epsfig{file=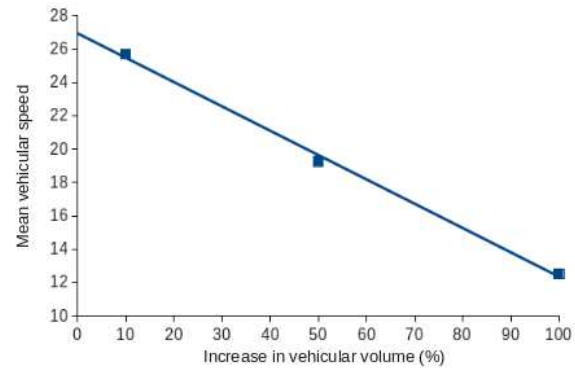, width=3in}
\caption{The respective mean~$\Sigma$ of ID3 when~$\mathcal{V}$ was increased from 10\% to 100\%. The line represents the regression with $\Sigma = -0.15 \times \mathcal{V} + 26.93$ ($R^2>0.99$).}\label{fig:regress2}
\end{figure}

Figures~\ref{fig:regress1} and~\ref{fig:regress2} show the respective effect of increasing~$\mathcal{V}$ from 10\% to 100\% on~$\Delta$ and~$\Sigma$ while under ID3. As~$\mathcal{V}$  is increased one percent at a time, $\Delta$~increases linearly with $R^2=0.98$ at a rate of 1.03~$s$ while $\Sigma$~decreases linearly with $R^2>0.99$ at a rate of 0.15~$km/h$. At this linear rate of decrease, all vehicles will come to a halt (i.e., $\Sigma=0$) at $\mathcal{V}\approx 180\%$.

\section{Conclusion}\label{sec:conclude}

In this paper, we presented a multi-agent-based microsimulation of a real-world vehicular traffic based on the existing psycho-physical car following model for simulating the movement of vehicles along the lane~\citep{fritzsche94,gipps81} and a rule-based algorithm for simulating the lane-changing behaviors of the vehicles~~\citep{alshihabi03,VISSIM90}. We simulated a real-world infrastructure involving six routes merging into a three-road fork via an unsignalized roundabout. This roundabout is the BR located in Taguig City, Metro Manila. From the observed distribution of vehicles per route and per vehicular type, we simulated the vehicular traffic of the current situation and collected the~$\tau$ metric. Based on the ANOVA, we found out that~$\tau_o$ and~$\tau_s$ are not statistically different from each other based on the F~statistics. This means that our simulation model can represent the real-world. We then simulated three proposed infrastructure developments (ID1, ID2, and ID3) and found out that ID3 offers the least~$\Delta$ and at the same time the fastest $\Sigma$. We increased the $\mathcal{V}$ from 10\% to 100\% under ID3 and found out that $\Delta$ increases linearly with $\mathcal{V}$, while $\Sigma$ decreases linearly. Based on these results, we provide the following conclusions:
\begin{enumerate}
\item Our microsimulation model can stastically represent the real-world;
\item The best ID for BR is ID3; and
\item Under ID3, $\mathcal{V}$ has a positive linear effect on~$\Delta$ and a negative linear effect on~$\Sigma$.
\end{enumerate}

\section{Recommendations}\label{sec:recommend}
We recommend to the stakeholders within BR, such as the City Government of Taguig through its TMO, as well as the Engineering Section of the MMDA, the residents of Lower Bicutan, the public and private employees within BR such as the DOST, the TUP-T, and the PUP-T, and the private sector, particularly SM Corporation, to help call for the implementation of infrastructure developments described in ID3. We caution, however, that we may be able to find a better vehicular flow if hybrid improvements of ID1 through ID3 are considered. We further caution that it may be cheaper to implement signalized traffic schemes together with the ID. As we have mentioned before, we have already reported our findings with various traffic schemes elsewhere~\cite{tataro13}, while we are currently working~\cite{arada14} on the interactive effects of ID and traffic schemes to both~$\Delta$ and~$\Sigma$.

\section{Future Work}
Several scientific research efforts have already forked out from this initial investigation. These are:
\begin{enumerate}
\item Studying the effects of installing railings in busy sidewalks;
\item Pedestrian dynamics for a very, very large academic building beyond its designed full capacity;
\item Evaluation of evacuation plans for Tsunami-vulnerable shoreline communities;
\item Designing re-routable and adaptive traffic plans for time-responsive traffic schemes; and
\item Incorporating the ``greedy'' behavior of Jeepney and tricycle drivers in the car-following model.
\end{enumerate}

\section{Acknowledgment}

This research effort is funded by the DOST-SEI Graduate Scholarship Program with M.G.~Arada and M.F.~Tataro as scholar-grantees taking the Master of Information Technology graduate program in the \UPLB\ under the research supervision of J.P.~Pabico. We thank the PTV Group~\citep{PTV} for allowing us to use their visual simulation framework free of charge for academic and research purposes. 

The following are the respective contributions of the authors: 
\begin{enumerate}
\item M.G.~Arada implemented the computational solution, conducted the real-world observation, the computational experiments and the statistical analyses, and prepared the final manuscript; 
\item M.F.~Tataro implemented the computational solution and conducted the real-world observation and computational experiments; and 
\item J.P.~Pabico formulated the computational solution to the problem, conducted and interpreted the statistical analyses, and edited the final manuscript. 
\end{enumerate}
All authors declare no conflict of interest.

\bibliography{paper1}
\bibliographystyle{plainnat}

\end{document}